\documentclass[11pt]{article}
\usepackage[english]{babel}
\usepackage{amssymb,amsmath,graphicx}
\usepackage{amsthm}
\usepackage[round,authoryear]{natbib}
\usepackage{setspace}
\usepackage{algorithmic}
\usepackage{algorithm}
\usepackage[table]{xcolor}
\usepackage{multirow}
\usepackage{natbib}
\usepackage{microtype}
\usepackage[caption=false]{subfig}
\usepackage{url}

 \bibpunct[, ]{(}{)}{,}{a}{}{,}%
 \def\newblock{\ }%

\title{Benchmark Instances and Branch-and-Cut Algorithm for the Hashiwokakero Puzzle}

\author{Leandro C.~Coelho, Gilbert Laporte, Arinei Lindbeck, Thibaut Vidal}

\begin{document}

\begin{center}

\begin{LARGE}
Benchmark Instances and Branch-and-Cut Algorithm for the Hashiwokakero Puzzle
\end{LARGE}

\vspace*{01cm}

\textbf{Leandro C.~Coelho$^{1,2}$, Gilbert Laporte$^{1,3}$, Arinei Lindbeck$^4$, Thibaut Vidal$^5$} \\
$^1$ Interuniversity Research Center on Enterprise Networks, Logistics \linebreak and Transportation (CIRRELT) \\
$^2$ Canada Research Chair in Integrated Logistics, Universit\'e Laval, Canada \\
$^3$ Canada Research Chair in Distribution Management, HEC Montr\'{e}al, Canada \\
$^4$ Universidade Federal do Paran\'a, Brazil \\
$^5$ Departamento de Inform\'{a}tica, Pontif\'{i}cia Universidade Cat\'{o}lica do Rio de Janeiro (PUC-Rio) \\

\vspace*{0.4cm}

Technical Report -- PUC-Rio -- May 2019

\vspace*{0.8cm}

\end{center}

%
%
%
%
%
%
%
%

\noindent
\textbf{Abstract.} Hashiwokakero, or simply Hashi, is a Japanese single-player puzzle played on a rectangular grid with no standard size. Some cells of the grid contain a circle, called \textit{island}, with a number inside it ranging from one to eight. The remaining positions of the grid are empty. The player must connect all of the islands by drawing a series of horizontal or vertical bridges between them, respecting a series of rules: the number of bridges incident to an island equals the number indicated in the circle, at most two bridges are incident to any side of an island, bridges cannot cross each other or pass through  islands, and each island must eventually be  reachable from any other island. In this paper, we present some complexity results and relationships between Hashi and well-known graph theory problems. We give a formulation of the problem by means of an integer linear mathematical programming model, and apply a branch-and-cut algorithm to solve the model in which connectivity constraints are dynamically generated. We also develop a puzzle generator. Our experiments on 1440 Hashi puzzles show that the algorithm can consistently solve hard puzzles with up to 400 islands.

\vspace*{0.2cm}


\vspace*{0.5cm}

\thispagestyle{empty}
\pagenumbering{arabic}

\section{Introduction}

\begin{figure}[hbtp]
        \centering
                \includegraphics[width=0.7\textwidth]{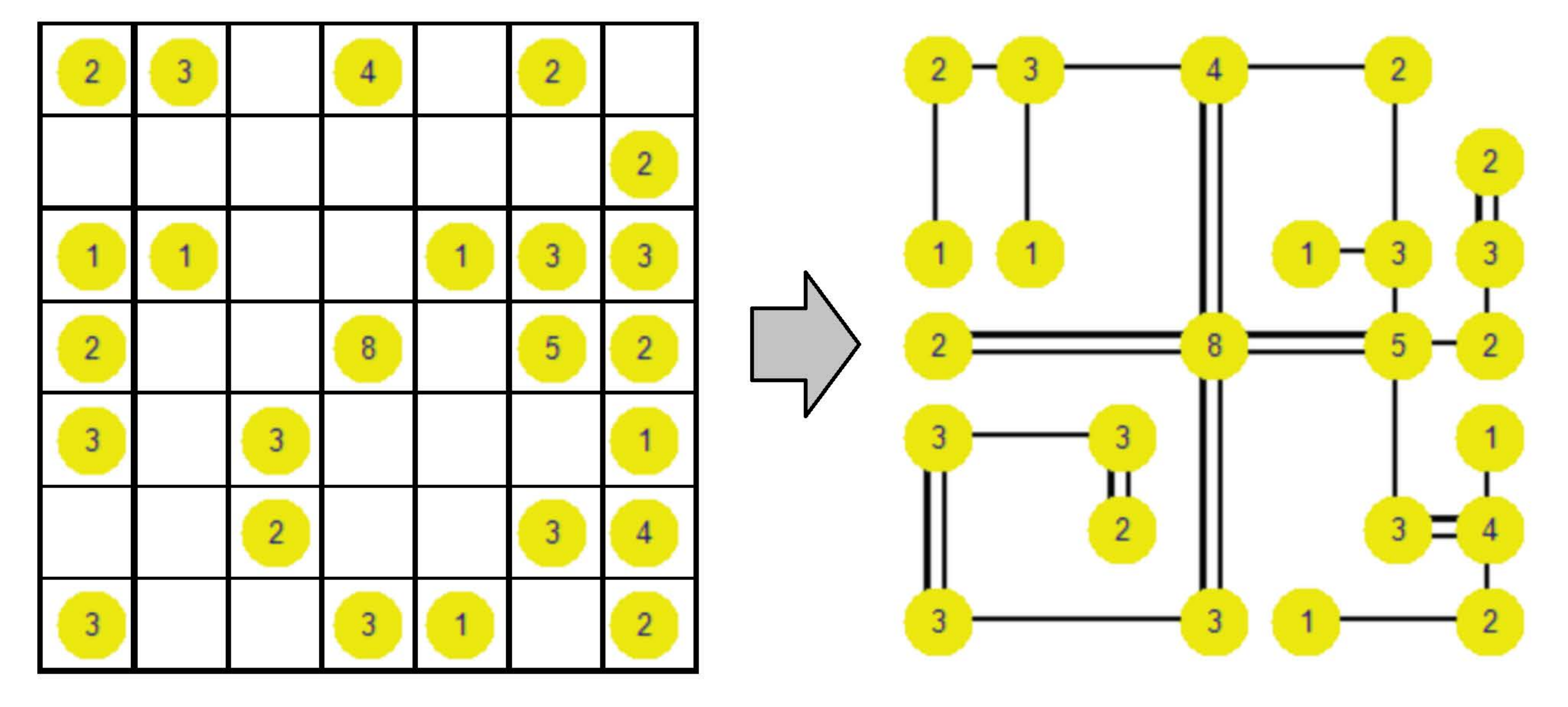}
        \caption{A Hashi puzzle (left) and a feasible solution (right)}\label{quick}
\end{figure}

Hashiwokakero, or simply Hashi, is a Japanese single-player puzzle played on a rectangular grid with no standard size. Some cells of the grid contain a circle, called \textit{island}, with a number inside it ranging from one to eight, and the number of islands is denoted by $n$. The remaining positions of the grid are empty. The player must connect all  the islands by drawing bridges between them. For this reason, the game is often referred to as \textit{building bridges}. The solution to the puzzle must respect the following rules:

\begin{enumerate}
\item the bridges must begin and end at distinct islands;
\item they must not cross any other bridges or islands;
\item they may only run horizontally or vertically;
\item at most two bridges may connect any pair of islands;
\item the number of bridges connected to each island must be equal to the number inscribed in the circle;
\item each island must be reachable from any other island.
\end{enumerate}
Figure \ref{quick} depicts a 7$\times$7 puzzle with $n = 24$ islands along with a feasible solution.

Unlike the Sudoku (see, e.g., \citealt{Coelho2013b}) the literature on Hashi is rather scarce. Perhaps the most important contribution to the Hashi literature is that of \citet{Andersson2009} who proved the problem to be NP-complete by reduction from a Hamiltonian circuit in unit-distance graphs.
Therefore, like all other NP-complete problems, such as the Hamiltonian cycle problem~\citep{Lawler1985}, the Hamiltonian path problem \citep{Garey1976}, the maximum cut problem \citep{Gavril1977}, the map coloring problem \citep{Dahl1987}, finding cliques of given sizes \citep{Karp1972}, the Hashi problem can be reduced to a Boolean satisfiability problem. 

We note that the complexity proof of \cite{Andersson2009} is fundamentally based on rule~6 (connectivity). Nevertheless, even without rule~6, we show that the problem remains NP-complete via a different reduction. Indeed, the problem of \emph{reconstructing disjoint sets of orthogonal segments} (RDOS) can be reduced to a Hashi instance in which all islands have value~one, and therefore where no double bridges are needed. RDOS has been proven to be NP-complete using an elegant reduction from 3-SAT in \citet{Rendl1993}. Finally the problem without the connectivity and no-crossing rules (2 and 6) containing only islands of value one is solvable by an $O(n \log n)$ polynomial algorithm.
 
Beyond complexity results, some algorithms have been presented.
\citet{Malik2012} combined heuristics operators and a backtracking algorithm to find feasible solutions. The operators are based on the intuitive decisions that a player would make when solving the game. When no such operators can be feasibly applied, a backtracking procedure takes place to explore other solutions. Some other papers have considered Hashi as part of a larger framework. \citet{Golan2011} shows that any Hashi puzzle can be reduced to a minesweeper puzzle. While studying trees and graphs, \citet{Prosser} proposed two new constraints for modeling trees, and used the Hashi puzzle as an illustration of their technique. Finally, \citet{Brain2009} used Hashi to illustrate how to efficiently implement algorithms based on answer set programming, a declarative programming paradigm used to model difficult search problems.

In this paper, we introduce new benchmark instances as well as a mathematical programming model and branch-and-cut algorithm for the Hashi puzzle. The remainder of this paper is organized as follows: Section \ref{model} describes the model and algorithm; Section \ref{instance} develops a puzzle generator and Section \ref{compu} analyzes the performance of our algorithm on a large set of instances; conclusions follow in Section \ref{conc}.

\section{Mathematical model and branch-and-cut algorithm}\label{model}

The Hashi puzzle can be defined on an undirected graph $\mathcal{G} = (\mathcal{V}, \mathcal{E})$, where $\mathcal{V}$ is the set of vertices representing islands. Let $d_i$ be the number of bridges to be constructed from island $i$, and $|V| = n$. Let $\delta(i)$ be the set of vertices adjacent to vertex $i$ either horizontally or vertically. Let $\mathcal{E}$ be the set of all edges connecting two adjacent vertices of $\mathcal{V}$. By convention, if $(i, j) \in \mathcal{E}$, then $i<j$. Let $\Delta$ be the set of intersecting edge pairs $\{(i,j), (k,l)\} \in \mathcal{E}$.
We model Hashi as an integer linear program which admits a solution if and only if the Hashi puzzle is feasible. For $(i,j) \in \mathcal{E}$, our model uses binary variables $y_{ij}$ indicating whether two adjacent vertices $i$ and~$j$ are connected by at least one bridge in the solution, and integer variables $x_{ij}$ indicating the number of bridges between $i$ and $j$. The formulation is then:
\begin{align}
\label{degree}
 \sum\limits_{i<k, i \in \delta(k)} x_{ik} + \sum\limits_{j > k, j \in \delta(k)} x_{kj} &= d_k && k \in \mathcal{V} \\
\label{link}
 y_{ij} \leq x_{ij} &\leq 2y_{ij} && (i,j) \in \mathcal{E}\\
\label{crossing2}
 y_{ij} + y_{kl} &\leq 1 && \{(i,j), (k,l)\} \in \Delta\\
\label{SEC}
 \sum\limits_{\substack{i \in S, j \in \mathcal{V} \backslash S \\ \textrm{or}~j \in S, i \in \mathcal{V} \backslash S}} y_{ij} &\geq 1 && S \subset \mathcal{V}, 1 \leq |S| \leq n-1\\
\label{bounds1}
 x_{ij} &\in \{0, 1, 2\} && (i, j) \in \mathcal{E}\\
\label{bounds2}
 y_{ij} &\in \{0, 1\} &&  (i, j) \in \mathcal{E}.
\end{align}

Constraints \eqref{degree} force the presence of $d_k$ bridges for each vertex $k$. According to constraints~\eqref{link}, at most two bridges can exist between any two connected vertices. These constraints also ensure consistency between the $x_{ij}$ and $y_{ij}$ variables. Constraints~\eqref{crossing2} prohibit intersecting bridges, and constraints \eqref{SEC} are strong connectivity constraints, enforcing the solution to be connected, as in the traveling salesman problem \citep{Dantzig1954}. Constraints \eqref{bounds1} and \eqref{bounds2} define the domains of the variables.

This formulation can be strengthened by adding a valid inequality which exploits the fact that the graph induced by the positive $y_{ij}$ variables must contain a spanning tree. It is called ``weak connectivity constraint'' and was found to be helpful in an algorithm in which the strong connectivity constraints \eqref{SEC} are initially relaxed:
\begin{equation}\label{noCycle}
\sum\limits_{(i,j) \in \mathcal{E}} y_{ij} \geq n - 1.
\end{equation}

To solve the problem, we use a branch-and-cut algorithm which initially relaxes constraints~\eqref{SEC}, and then detects (separates) and reintroduces the offending constraints each time an integer solution of the resulting branch-and-bound tree is found to be infeasible. 

\begin{figure}[hbtp]
        \centering
        \subfloat[Hashi puzzle]{
                \includegraphics[width=0.3\textwidth]{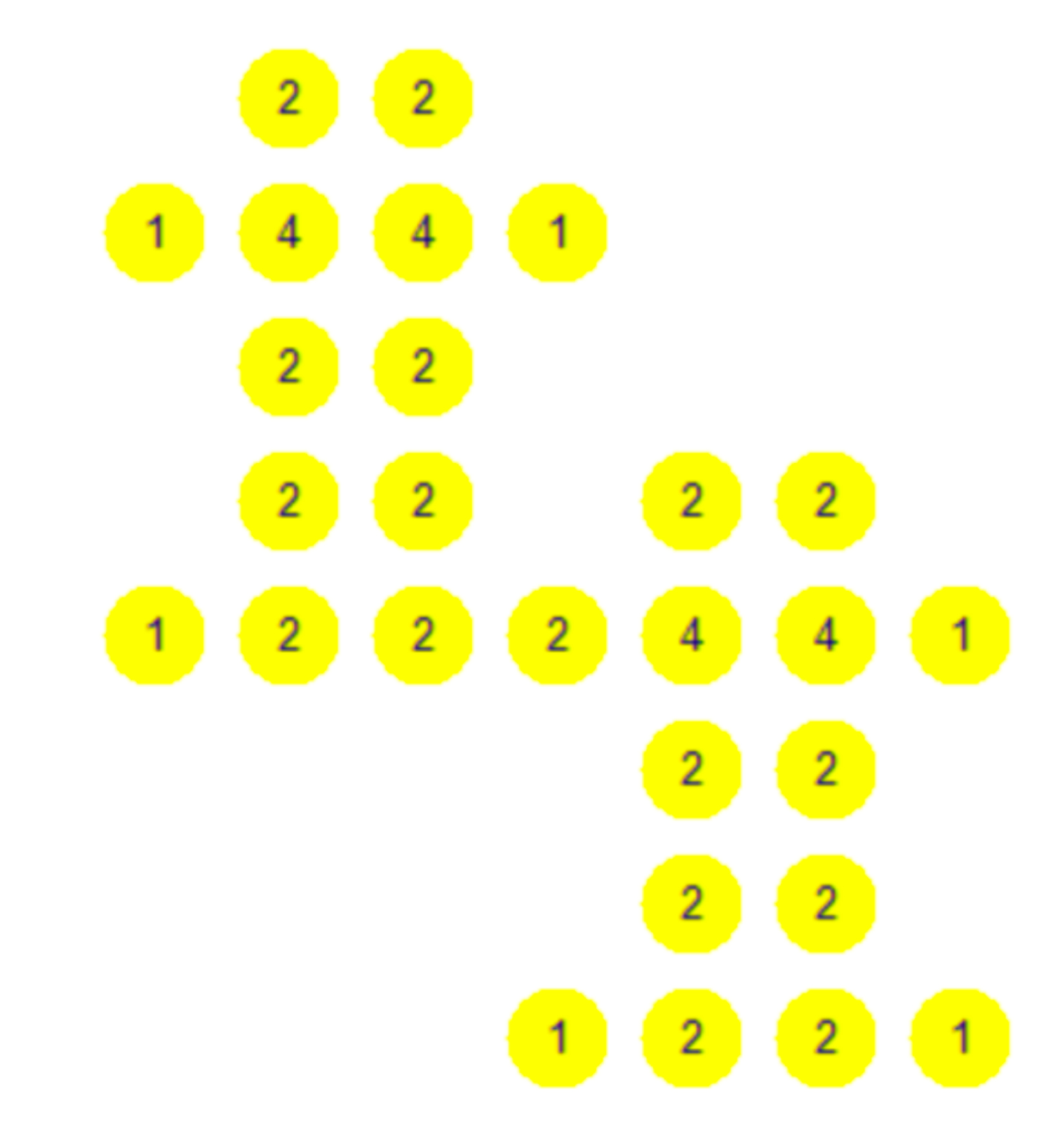}
                \label{slow1}
        }%
        
        \subfloat[Intermediate steps: dynamic introduction of the strong connectivity constraints]{
                \includegraphics[width=0.9\textwidth]{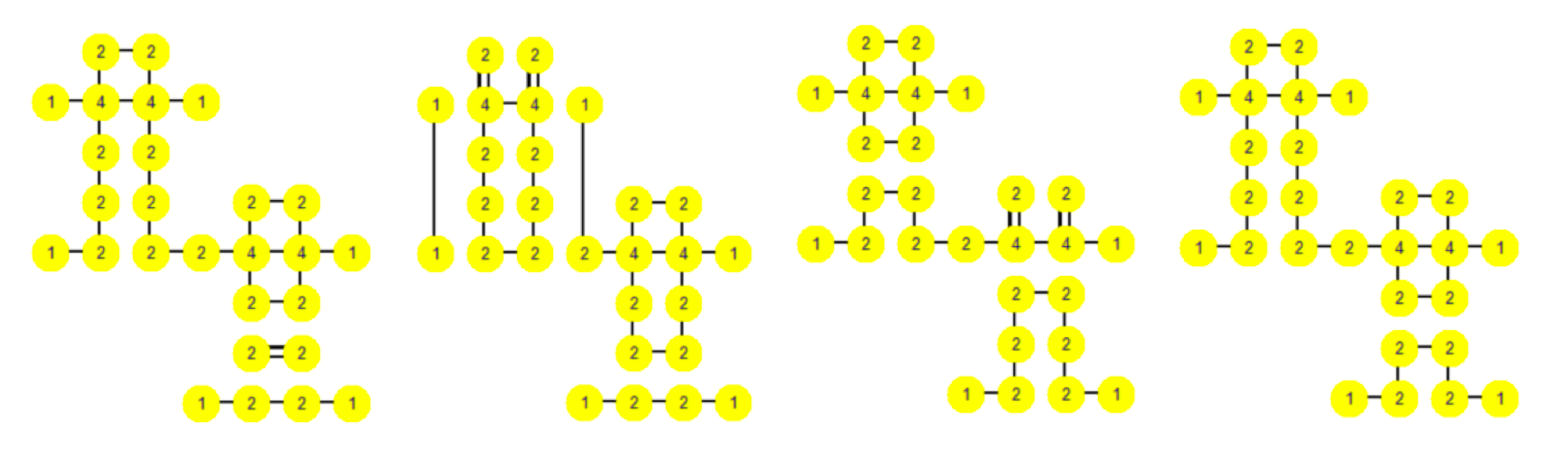}
                \label{slow2}
        }

        \subfloat[The completed Hashi]{
                \includegraphics[width=0.3\textwidth]{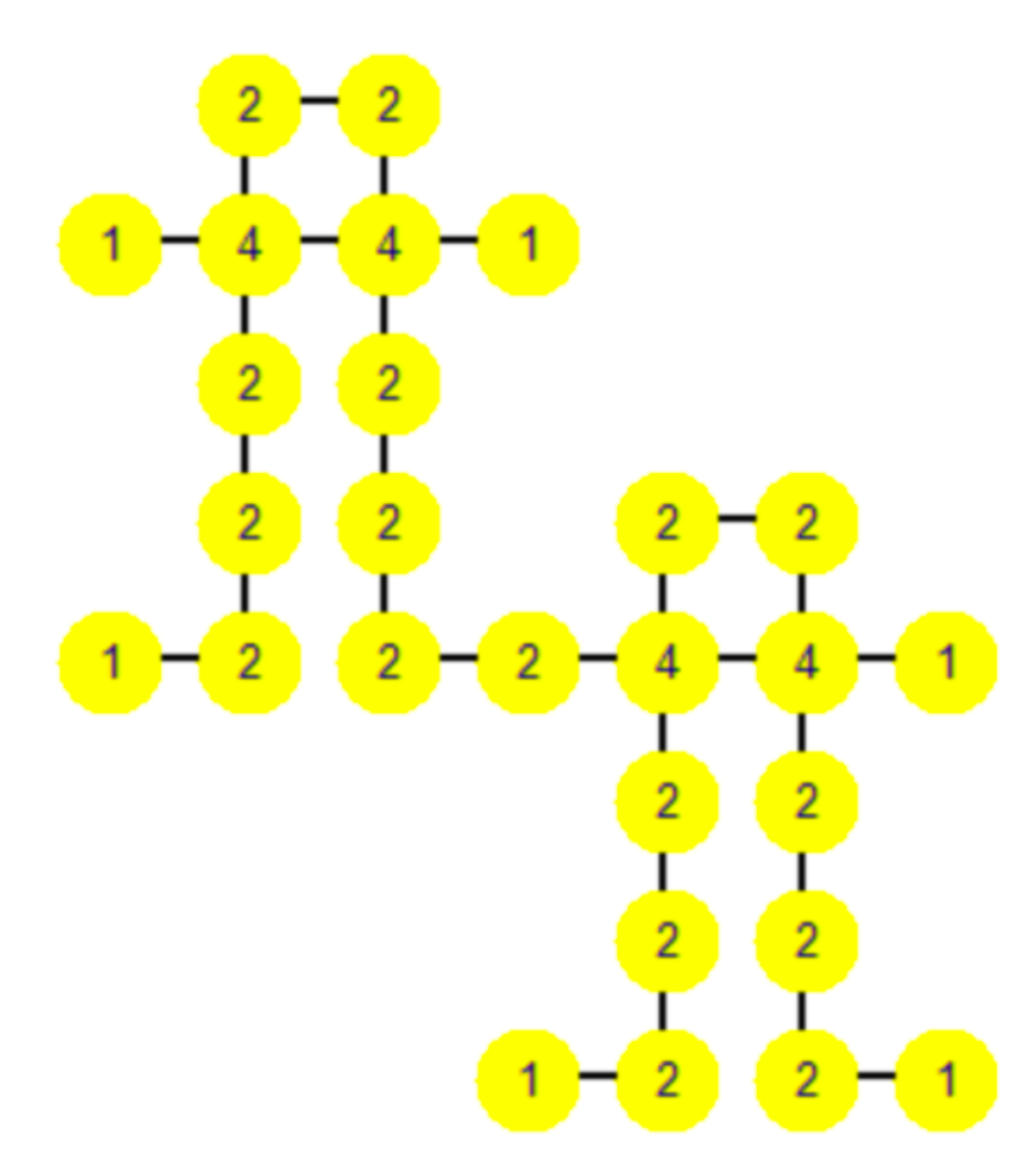}
                \label{slow3}
        }
        \caption{Example of a difficult Hashi requiring four strong connectivity constraints}\label{slow}
\end{figure}

On the example of Figure \ref{quick} it was possible to find a feasible solution without need for additional strong connectivity constraint. In contrast, Figure \ref{slow} shows a new puzzle which required four added strong connectivity constraints, along with their partial solutions obtained with our algorithm. It is important to note that without the weak connectivity constraint \eqref{noCycle}, 87 strong connectivity constraints would have been needed.

\section{Puzzle generator}
\label{instance}

This section describes an algorithm designed to generate large scale Hashi puzzles which will be used for our experimental analyzes. The generator receives as input the desired number of islands $n$, the dimensions of the grid $d_1 \times d_2$, a parameter $\alpha \in [0\%,100\%]$ which influences the number of cycles and the connectivity of the solutions, and a parameter $\beta \in [0\%,100\%]$ which influences the number of double-bridges of the solutions. The generator is made up of four steps:
\begin{itemize}
\item
\textbf{Step 1 -- Placement of the islands.} A first island is placed in a random grid location. Then, the algorithm iteratively selects a random existing island, a random direction (top, bottom, left or right), and a random position in this direction (without crossing an existing edge) if possible to add a new island and edge. This process is repeated $n-1$ times.

\item
\textbf{Step 2 -- Creating cycles.} At the end of Step 1, there are $n$ islands connected by $n-1$ edges. To avoid studying particular cases of Hashi puzzles that admit trees as solutions,  the tree is augmented with additional edges, therefore creating cycles. Iteratively, the algorithm randomly selects two islands which can be connected by a horizontal or vertical edge without crossing existing ones, and connects them with one additional edge. This process is repeated $\lfloor \alpha n \rfloor$ times, leading to $n$ islands connected with $n-1+\lfloor \alpha n \rfloor$ edges.

\item
\textbf{Step 3 -- Creating double edges.} To favor the possible use of double edges in solutions, 
the algorithm iteratively considers each edge, and transforms this edge into a double-edge with probability $\beta$.

\item
\textbf{Step 4 -- Adjacency count and final puzzle.} Finally, the number of edges adjacent to each island is counted and marked. All edges are erased, and the puzzle is returned.
\end{itemize}

By design, all puzzles are known to be feasible since a feasible solution exists at the end of Step 3 before erasing the edge information. Also note that this solution may not be unique.

\section{Computational experiments}
\label{compu}

We have implemented our algorithms in Visual Basic and used Gurobi 8.0 to build the branch-and-cut algorithm. To generate a representative set of instances, we considered four possible instance size values, considering a number of islands $n \in \{100,200,300,400\}$ on boards of dimensions $d_1 \times d_2 \in \{16\times16,24\times24,28\times28,33\times33\}$ respectively. We varied the level of connectivity by selecting $\alpha \in \{0\%,5\%,10\%,15\%\}$ as well as the level of double-edges by considering $\beta \in \{0.25,0.5,0.75\}$. For each parameters combination (instance group), we generated 30 random instances to increase the statistical strength of our analyzes, leading to a total of $30 \times 4 \times 4 \times 3 = 1440$ Hashi instances divided into 48 groups. These instances can be accessed at \url{https://w1.cirrelt.ca/~vidalt/en/research-data.html}. We ran the branch-and-cut algorithm on each instance, and repeated the same experiment without the weak connectivity constraint \eqref{noCycle} to investigate its influence on the search performance.

Tables~\ref{noYes} and~\ref{noNo} present the results of our experiments with and without the weak connectivity constraint. The first set of columns reports the average CPU time needed to solve each instance of the group, and the second set of columns reports the average number of strong connectivity constraints that were dynamically generated during the solution process.

\begin{table}[hbpt]
\centering
\caption{Performance of the branch-and-cut algorithm with the weak connectivity constraint}\label{noYes}
\renewcommand{\arraystretch}{1.15}
\setlength\tabcolsep{8pt}
\scalebox{0.8}{
\begin{tabular}  { |cc| ccccc| ccccc|}
\hline
&& \multicolumn{5}{c|}{Time (s)} & \multicolumn{5}{c|}{Number of strong connectivity constraints} \\
n & $\beta$ & $\alpha=0\%$ & $5\%$ &   $10\%$ &  $15\%$ & Avg. & $\alpha=0\%$ & $5\%$ &   $10\%$ &  $15\%$ & Avg.  \\
\hline
\multirow{4}{*}{100}&0.25&0.05&0.04&0.04&0.03&0.04&1.10&0.70&0.57&0.40&0.69\\
&0.5&0.11&0.10&0.11&0.09&0.10&3.33&3.00&3.13&2.40&2.97\\
&0.75&0.22&0.39&0.39&0.36&0.34&5.93&11.77&12.03&11.07&10.20\\
&\textbf{Avg.}&\textbf{0.13}&\textbf{0.18}&\textbf{0.18}&\textbf{0.16}&\textbf{0.16}&\textbf{3.46}&\textbf{5.16}&\textbf{5.24}&\textbf{4.62} &\textbf{4.62}\\
\hline
\multirow{4}{*}{200}&0.25&0.16&0.15&0.13&0.13&0.14&2.10&1.63&1.13&0.90&1.44\\
&0.5&0.52&0.53&0.49&0.44&0.50&7.93&7.53&6.13&4.87&6.62\\
&0.75&2.13&2.87&3.15&2.91&2.76&31.13&41.40&45.50&38.80&39.21\\
&\textbf{Avg.}&\textbf{0.94}&\textbf{1.18}&\textbf{1.26}&\textbf{1.16}&\textbf{1.13}&\textbf{13.72}&\textbf{16.86}&\textbf{17.59}&\textbf{14.86} &\textbf{15.76}\\
\hline
\multirow{4}{*}{300}&0.25&0.43&0.38&0.40&0.35&0.39&3.47&2.23&2.43&1.33&2.37\\
&0.5&2.34&1.86&1.89&1.69&1.95&18.33&11.73&10.43&7.60&12.03\\
&0.75&15.38&19.35&18.14&13.66&16.63&90.30&107.93&182.17&66.83&111.81\\
&\textbf{Avg.}&\textbf{6.05}&\textbf{7.20}&\textbf{6.81}&\textbf{5.23}&\textbf{6.32}&\textbf{37.37}&\textbf{40.63}&\textbf{65.01}&\textbf{25.26} &\textbf{42.07}\\
\hline
\multirow{4}{*}{400}&0.25&0.92&0.85&0.73&0.80&0.83&4.40&3.50&2.00&2.10&3.00\\
&0.5&6.89&5.71&6.55&7.05&6.55&28.90&16.47&19.70&14.17&19.81\\
&0.75&76.65&152.18&100.68&84.23&103.43&161.80&238.77&159.90&130.20&172.67\\
&\textbf{Avg.}&\textbf{28.15}&\textbf{52.91}&\textbf{35.99}&\textbf{30.69}&\textbf{36.94}&\textbf{65.03}&\textbf{86.24}&\textbf{60.53}&\textbf{48.82} &\textbf{65.16}\\
\hline
\multicolumn {2} {|c|} {\textbf{Avg. All}}&\textbf{8.82}&\textbf{15.37}&\textbf{11.06}&\textbf{9.31}&\textbf{11.14}&\textbf{29.89}&\textbf{37.22}&\textbf{37.09}&\textbf{23.39}&\textbf{31.90}\\
\hline
\end{tabular}}
\end{table}

\begin{table}[hbpt]
\centering
\caption{Performance of the branch-and-cut algorithm without the weak connectivity constraint}\label{noNo}
\renewcommand{\arraystretch}{1.15}
\setlength\tabcolsep{7.3pt}
\scalebox{0.8}{
\begin{tabular}  { | cc| ccccc| ccccc|}
\hline
&& \multicolumn{5}{c|}{Time (s)} & \multicolumn{5}{c|}{Number of strong connectivity constraints} \\
n & $\beta$ & $\alpha=0\%$ & $5\%$ &   $10\%$ &  $15\%$ & Avg. & $\alpha=0\%$ & $5\%$ &   $10\%$ &  $15\%$ & Avg.  \\
\hline
\multirow{4}{*}{100}&0.25&0.05&0.04&0.03&0.03&0.04&1.30&0.80&0.53&0.43&0.77\\
&0.5&0.14&0.13&0.12&0.11&0.13&4.33&3.80&3.67&3.40&3.80\\
&0.75&0.80&0.64&0.47&0.45&0.59&24.57&19.20&13.77&12.73&17.57\\
&\textbf{Avg.}&\textbf{0.33}&\textbf{0.27}&\textbf{0.21}&\textbf{0.20}&\textbf{0.25}&\textbf{10.07}&\textbf{7.93}&\textbf{5.99}&\textbf{5.52} &\textbf{7.38}\\
\hline
\multirow{4}{*}{200}&0.25&0.15&0.12&0.10&0.11&0.12&2.23&1.57&0.93&1.00&1.43\\
&0.5&0.52&0.42&0.54&0.48&0.49&9.23&6.57&7.07&6.13&7.25\\
&0.75&4.33&3.82&3.53&3.24&3.73&73.13&60.67&52.47&50.47&59.18\\
&\textbf{Avg.}&\textbf{1.67}&\textbf{1.46}&\textbf{1.39}&\textbf{1.28}&\textbf{1.45}&\textbf{28.20}&\textbf{22.93}&\textbf{20.16}&\textbf{19.20} &\textbf{22.62}\\
\hline
\multirow{4}{*}{300}&0.25&0.40&0.32&0.35&0.29&0.34&4.27&2.53&2.60&1.43&2.71\\
&0.5&2.16&1.80&1.67&1.86&1.87&18.77&12.43&9.07&8.93&12.30\\
&0.75&24.38&24.41&19.69&15.31&20.95&145.93&162.57&105.07&75.37&122.23\\
&\textbf{Avg.}&\textbf{8.98}&\textbf{8.85}&\textbf{7.24}&\textbf{5.82}&\textbf{7.72}&\textbf{56.32}&\textbf{59.18}&\textbf{38.91}&\textbf{28.58} &\textbf{45.75}\\
\hline
\multirow{4}{*}{400}&0.25&0.76&0.72&0.60&0.58&0.66&4.87&3.43&1.90&2.00&3.05\\
&0.5&6.13&5.92&6.53&5.96&6.14&23.77&18.17&17.53&12.43&17.98\\
&0.75&136.33&151.34&138.03&99.21&131.23&320.63&531.90&460.63&301.40&403.64\\
&\textbf{Avg.}&\textbf{47.74}&\textbf{52.66}&\textbf{48.39}&\textbf{35.25}&\textbf{46.01}&\textbf{116.42}&\textbf{184.50}&\textbf{160.02}&\textbf{105.28} &\textbf{141.56}\\
\hline
\multicolumn {2} {|c|} {\textbf{Avg. All}}&\textbf{14.68}&\textbf{15.81}&\textbf{14.31}&\textbf{10.64}&\textbf{13.86}&\textbf{52.75}&\textbf{68.64}&\textbf{56.27}&\textbf{39.64}&\textbf{54.33}\\\hline
\end{tabular}}
\end{table}

As observed in Table \ref{noYes}, our complete branch-and-cut algorithm appears to be very efficient, with an average CPU time of 11.14 seconds and an average of 31.90 strong connectivity constraints per instance. The smallest instances with 100 islands are solved within a fraction of a second, whereas the largest instances with 400 islands are significantly more difficult and require on average 36.94 seconds and 65.16 cuts. The proportion of double bridges has a significant impact on the puzzle difficulty, as reflected by a significant increase of CPU time and number of cuts when $\beta$ increases.

The results of our branch-and-cut algorithm without the weak connectivity constraint, reported in Table \ref{noNo}, lead to similar conclusions regarding problem difficulty as a function of $n$ and $\beta$. These results also clearly demonstrate the usefulness of the weak connectivity constraint~\eqref{noCycle}. Without it, the average CPU time rises up by 24\% and the number of strong connectivity cuts generated through the search rises up by 70\%.

Finally, the effect of the parameter~$\alpha$ (impacting the number of cycles in the solutions) is less marked. This is likely due to a combination of two effects: on the one hand, increasing $\alpha$ leads to solutions with a larger number of edges and makes the puzzle more complicated; on the other hand, higher $\alpha$ values help respecting the connectivity constraints, an effect which is especially visible in the results of Table \ref{noNo}, when the weak connectivity constraint is deactivated.

\section{Conclusions}
\label{conc}

We have designed a set of benchmark instances for the Hashi puzzle and proposed 
a branch-and-cut algorithm. We have conducted sensitivity analyses on the impact of the main instance parameters and features of the solution method. Our experiments demonstrate the good performance of the algorithm, which solves all puzzles with up to 400 islands, and the contribution of the weak connectivity constraint in the search.

\section*{Acknowledgments}

We thank Vanessa Dias for suggesting  this research topic to us. This work was partly supported by the Canadian Natural Sciences and Engineering Research Council (NSERC) under grants  2014-05764 and 2015-01689, as well as CAPES, FAPERJ and CNPq in Brazil. This support is gratefully acknowledged.

\section*{References}

\end{document}